\documentclass[preprint,showpacs,preprintnumbers,amsmath,amssymb]{revtex4}

\usepackage{graphicx}
\usepackage{dcolumn} 
\usepackage{bm}

\begin{document}

\title{Transport properties of the layered Rh oxide K$_{0.49}$RhO$_2$
}

\author{S. Shibasaki}
\email{g01k0374@suou.waseda.jp}
\affiliation{Department of Applied Physics, Waseda University, 
Tokyo 169-8555, Japan}

\author{T. Nakano}
\affiliation{Department of Applied Physics, Waseda University, 
Tokyo 169-8555, Japan}

\author{I. Terasaki}
\affiliation{Department of Applied Physics, Waseda University, 
Tokyo 169-8555, Japan}

\author{K. Yubuta}
\affiliation{Institute for Materials Research, Tohoku University, 
Sendai 980-8577, Japan}

\author{T. Kajitani}
\affiliation{Department of Applied Physics, Graduate School of Engineering, Tohoku University, 
Sendai 980-8577, Japan}

\date{\today}

\begin{abstract}
We report measurements and analyses of resistivity, thermopower and Hall coefficient of single-crystalline samples of the layered Rh oxide K$_{0.49}$RhO$_2$.
The resistivity is proportional to the square of temperature up to 300~K, and the thermopower is proportional to temperature up to 140~K.
The Hall coefficient increases linearly with temperature above 100~K, which is ascribed to the triangular network of Rh in this compound.
The different transport properties between Na$_x$CoO$_2$ and K$_{0.49}$RhO$_2$ are discussed on the basis of the different band width between Co and Rh evaluated from the magnetotransport.
\end{abstract}


\maketitle

\section{Introduction}
Since the discovery of large thermopower with low resistivity in Na$_x$CoO$_2$\cite{NCO},
Co oxides with the CdI$_2$-type CoO$_2$ layer such as Ca-Co-O and Bi-Sr-Co-O have been studied as thermoelectric materials\cite{CCO1,CCO2,BSCO_poly,BSCO_tp,BSCO_single}.
These materials have large thermopower comparable to conventional thermoelectric materials, which is considered to be due to the degeneracy of spins and orbitals\cite{koshibae}.
In addition to the large thermopower, these Co oxides show many interesting properties.
For example, Na$_x$CoO$_2$ exhibits magnetic order($x\sim 0.75$), charge order($x\sim 0.5$), and even superconductivity($x\sim 0.35$) in its hydrated form\cite{NCO_mag,NCO_co,NCO_sc}.
Bi-Sr-Co-O compound shows pseudogap, ferromagnetism, and negative magnetoresistance\cite{BSCO_tp,BSCO_mag}.

\begin{figure}[t]
  \begin{center}
   \includegraphics[width=8cm,clip]{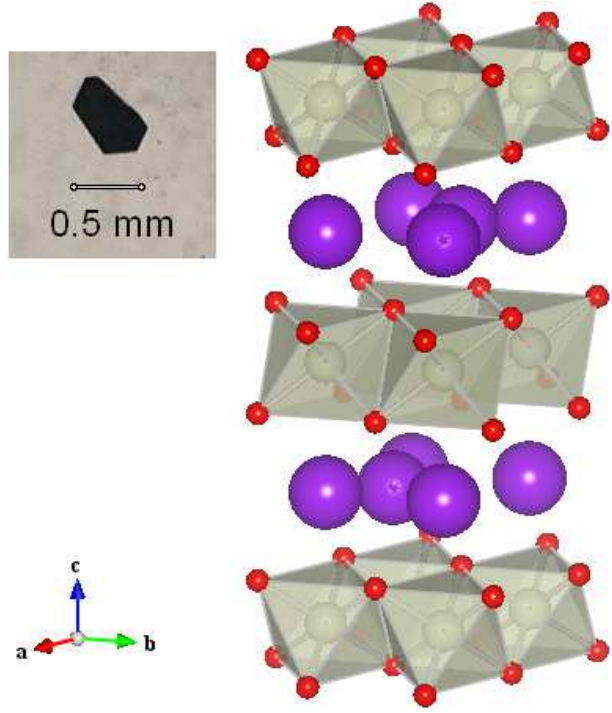}
   \caption{(Color online) Crystal structure of K$_{0.49}$RhO$_2$.
   Purple(dark gray), gray(light gray), and red(dark gray) represent K, Rh, and O atoms, respectively.
   This material crystallizes in the $\gamma$-Na$_x$CoO$_2$-type structure.
   A photographic image of a sample of K$_{0.49}$RhO$_2$ is shown at the left upper corner.
   }
   \label{fig1}
  \end{center}
\end{figure}

The layered Rh oxides with the CdI$_2$-type RhO$_2$ layer are found to have good thermoelectric properties similar to Co oxides\cite{BSRO,BBRO,BBRO_klein,SRO,CRO_kuriyama,CRO,BSRO_single}, 
which indicates that Rh ions in these Rh oxides have similar electronic states to Co ions in the layered Co oxides\cite{koshibae}.
Especially the Bi-Ba-Rh-O compound which is isomorphic to the Bi-Sr-Co-O compound shows pseudogap and negative magnetoresistance\cite{BBRO,BBRO_klein}.
We focus on K$_x$RhO$_2$ which can be a reference material to Na$_x$CoO$_2$.
Recently, we successfully prepared single-crystalline samples of K$_{0.49}$RhO$_2$\cite{KRO_yubuta}.
However, there are few reports on $A_x$RhO$_2$ ($A$ : alkali metals) and its hydrated form\cite{ARO,NRO,NRO_mag,NROH2O_res}, and no report on their thermoelectric properties.
Figure~\ref{fig1} shows the crystal structure of K$_{0.49}$RhO$_2$ and a photographic image of a sample\cite{VESTA}.
K$_{0.49}$RhO$_2$ crystallizes in the $\gamma$-Na$_x$CoO$_2$ type structure (space group $P6_3/mmc$) with the lattice parameters of $a=3.0647(4)$~$\textrm{\AA}$ and $c=13.600(2)$~$\textrm{\AA}$\cite{KRO_yubuta}.
The CdI$_2$-type RhO$_2$ layer and the K$_x$ layer are alternately stacked along the \textit{c}-axis direction.
Typical size of the sample is 0.5$\times$0.3$\times$0.05~mm$^3$.
In this paper, we report the transport properties of single-crystalline samples of K$_{0.49}$RhO$_2$ and compare them with those of other layered Co/Rh oxides.

\section{Experimental}
Single-crystalline samples of K$_{0.49}$RhO$_2$ were prepared by a flux-growth method.
Mixture of K$_2$CO$_3$ and Rh$_2$O$_3$ with a molar ratio of 25 : 1 was kept at 1373~K for 1~h, slowly cooled down to 1023~K at a rate of 2~K/h, and then cooled down to room temperature.
K$_2$CO$_3$ was used as a self-flux. As-grown crystals were removed from the products by washing with distilled water.
The crystals were characterized by electron probe micro-analysis and x-ray diffraction, and were identified to be isomorphic to $\gamma$-Na$_x$CoO$_2$\cite{KRO_yubuta}.

Resistivity and thermopower were measured by a conventional dc four-probe method and steady-state method, respectively.
Hall coefficient was measured with physical property measurement system(Quantum Design).
Current and temperature gradient were applied along the in-plane direction, and magnetic field was applied along the out-of-plane direction.

\section{Results and Discussion}
\begin{figure}[t]
  \begin{center}
   \includegraphics[width=8cm,clip]{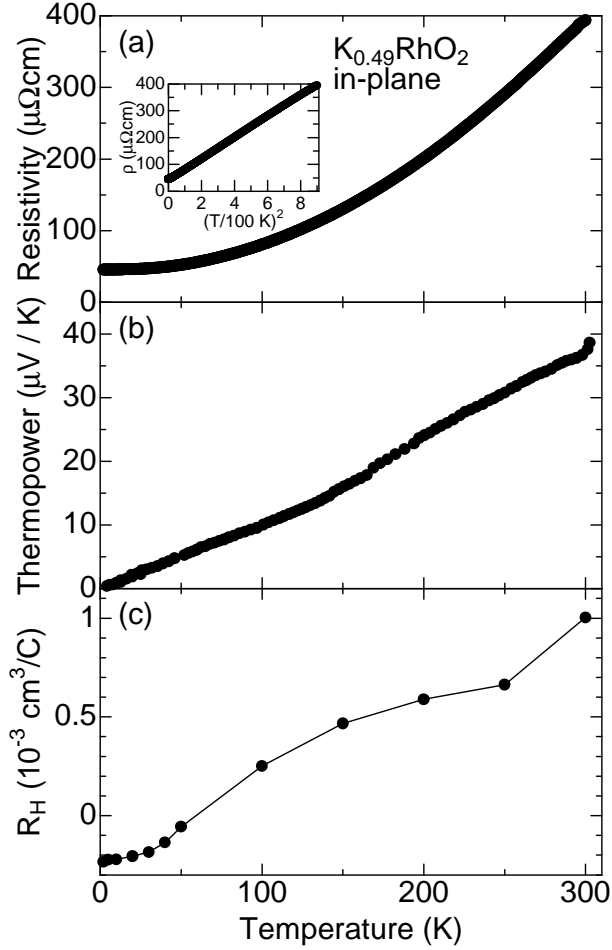}
   \caption{(a) In-plane resistivity, (b) thermopower, and (c) Hall coefficient of K$_{0.49}$RhO$_2$. Two insets show $T^2$-behavior of the resistivity.}
   \label{fig2}
  \end{center}
\end{figure}

Figure~\ref{fig2} shows the temperature dependence of the in-plane resistivity ($\rho$), thermopower ($S$), and Hall coefficient ($R_\textrm{H}$) of K$_{0.49}$RhO$_2$.
$\rho$ shows metallic behavior with the magnitude of 400~$\mu\Omega$cm at 300~K, which is twice as large as that of Na$_x$CoO$_2$\cite{NCO}.
This fairly low resistivity suggests a high mobility of the CdI$_2$-type RhO$_2$ layer like isomorphic CoO$_2$ layer\cite{BSRO,BBRO,BBRO_klein,SRO,CRO_kuriyama,CRO,BSRO_single}.
$\rho$ shows Fermi-liquid behavior ($\rho$=$AT^2+B$, where $T$ is temperature) up to 300~K, where $A$ and $B$ are $4.0\times 10^{-9}$~$\Omega$cm/$\textrm{K}^2$ and $4.2\times 10^{-5}$~$\Omega$cm, respectively.
Compared to the heavy-fermion systems, it is unusual that the $T^2$ power law is observed up to 300~K, but 
one can also see the $T^2$ power law up to high temperatures in doped SrTiO$_3$\cite{STO} and doped CuRhO$_2$\cite{Maignan}.
This behavior is completely different from Na$_x$CoO$_2$ whose resistivity is not proportional to $T^2$\cite{NCO_co,NCO_tp}. 
This suggests that the electronic correlation in K$_{0.49}$RhO$_2$ is weaker than that in Na$_x$CoO$_2$, although the $T^2$ law indicates significant correlation.
Indeed, $\rho$ of Na$_x$CoO$_2$ is proportional to $T^2$ under the magnetic field\cite{NCO_mr}, 
and we expect that the spin fluctuations are weaker in K$_{0.49}$RhO$_2$ than in Na$_x$CoO$_2$, because the magnetic field suppresses the spin fluctuations.

As shown in Fig.~\ref{fig2}(b), the magnitude of $S$ is 40~$\mu$V/K at 300~K, which is half of that of Na$_x$CoO$_2$\cite{NCO}.
$S$ is almost proportional to $T$ up to 300~K with a kink at 140~K, while $S$ of Na$_{0.71}$CoO$_2$ sample is almost $T$-independent around 300~K\cite{NCO_tp}.
Kuroki and Arita suggest that a ``pudding-mold'' band can explain the transport properties of Na$_x$CoO$_2$\cite{NCO_kuroki}.
Compared to $S$ of Na$_x$CoO$_2$, correlation in K$_{0.49}$RhO$_2$ is weak because $S$ increases almost linearly with $T$ up to 300~K(see eq.~(\ref{eq1})).
However, the kink around 140~K is not seen in the resistivity.
In SrRhO$_3$ and Sr$_3$Rh$_2$O$_7$, the thermopower shows a similar kink near 140~K\cite{SRO_s}.
Yamaura \textit{et al.} speculated that the kink in the thermopower may be structurally related.
This suggests that Rh oxides commonly show this thermopower anomaly.
As for the thermoelectric materials, the largest power factor $S^2/\rho$ of K$_{0.49}$RhO$_2$ is 4.0~$\mu$W/cmK$^2$ at 300~K.
Although this value is 10 times smaller than that of Na$_x$CoO$_2$, it is largest among the layered Rh oxides\cite{BSRO,BBRO,BBRO_klein,SRO,CRO_kuriyama,CRO,BSRO_single}.

As shown in Fig.~\ref{fig2}(c), the magnitude of $R_\textrm{H}$ is 1.0$\times 10^{-3}$~cm$^3$/C at 300~K, which is comparable to that of Na$_x$CoO$_2$ at 300~K\cite{NCO_hall}.
Temperature dependence of $R_\textrm{H}$ seems similar to those of other layered Co/Rh oxides\cite{NCO_co,BSRO_single,BSCO_hall,NCO_hall,NCPO,CCO_hall,SCO_hall,CNCO_hall} which are not constant like conventional metals. 

Let us discuss the transport properties of K$_{0.49}$RhO$_2$ quantitatively.
We evaluate the Fermi energy of this compound from the slope in the thermopower below 140~K 
using the following equation\cite{thermopowerlt}
\begin{equation}
 S=-\frac{\pi^2}{2e}\frac{k_\textrm{B}^2T}{E_\textrm{F}},
 \label{eq1}
\end{equation}
where $e$, $k_{\textrm{B}}$, and $E_{\textrm{F}}$ represent the elementary charge, the Boltzmann constant, and the Fermi energy, respectively.
The evaluated $E_{\textrm{F}}$ is 4200~K which is much higher than 140~K, and this is consistent with $T$-linear behavior of $S$.
From the crystal structure of K$_{0.49}$RhO$_2$, this material is regarded as a two-dimensional material.
$E_\textrm{F}$ for a two-dimensional system is written as\cite{fermienergy2d}
\begin{equation}
 E_{\textrm{F}}=\hbar^2\pi d_\textrm{c} \frac{n}{m^*},
 \label{eq2}
\end{equation}
where $\hbar$, $d_\textrm{c}$, $n$, and $m^*$ represent the Planck constant, the interlayer distance, the carrier concentration, and the effective mass, respectively.
Here, we adopt $d_\textrm{c}=c/2=6.8$~$\textrm{\AA}$ from crystallographic analysis\cite{KRO_yubuta}.
If we adopt $n$ evaluated from the composition, we will estimate the effective mass from eq.~(\ref{eq2}) to 80 times larger than the bare mass of electrons.
This value mostly comes from the narrow $t_{2g}$ bands.
The band calculations of NaCoO$_2$ and NaRhO$_2$ revealed that the electronic states are similar\cite{BBRO}.
We further note that the calculated band width of NaRhO$_2$ is comparable to that of CuRhO$_2$\cite{Maignan}.
Thus the actual band widths of Rh oxides can be regarded comparable to those of Co oxides.
The mobility is evaluated from $\rho$ to be 3.4~cm$^2$/V$\cdot$s at 300~K, which is comparable to that of Na$_x$CoO$_2$\cite{NCO}.

The thermopower and the electronic specific heat coefficient have the relation\cite{sgscale,misfitscale}
\begin{equation}
 \frac{S}{T}\frac{N_\textrm{\scriptsize{AV}}e}{\gamma}=\textrm{const.},
 \label{eqst}
\end{equation}
where $N_\textrm{\scriptsize{AV}}$ represents Avogadro number and constant value is of the order of 1, and we adopt constant value of 1.
This relation indicates that K$_{0.49}$RhO$_2$ has smaller electron specific heat coefficient $\gamma$ than Na$_x$CoO$_2$, because ot the smaller thermopower slope, 
which means that the correlation is weaker in Rh oxides.
We can also estimate $\gamma$ from 
Kadowaki-Woods relation($A/\gamma^2=1.0\times10^{-5}$~$\mu\Omega$cm(mol$\cdot$K/mJ)$^2$) using the $T^2$ coefficient in $\rho$\cite{kwrelation,kwrelation2}.
The estimated values are of the order of $10^{-2}$~J/mol$\cdot$K, and reasonably agree.

Now we discuss the $T$-dependent $R_\textrm{H}$ of K$_{0.49}$RhO$_2$.
First possibility is that the carrier concentration changes with temperature.
$R_\textrm{H}$ of K$_{0.49}$RhO$_2$ increases with temperature, and changes its sign around 50~K.
In the Drude model (free carrier in one band), the change in the sign of $R_\textrm{H}$ is unphysical.
Second possibility is that there are two kinds of carriers.
The sign of $R_\textrm{H}$ is negative below 50~K, which means that electrons are majority carrier as long as mobilities for electrons and holes are equal.
Then the change in the sign of $R_\textrm{H}$ around 50~K indicates that holes are dominant at high temperature.
This behavior is observed in Sr$_2$RuO$_4$.
Positive $S$ with negative $R_\textrm{H}$ in Sr$_2$RuO$_4$ is ascribed to the ratio of the densities between holes and electrons\cite{SRO_rh1,SRO_tp}, 
and the $dS/dT$ has an anomaly around 20~K where $R_\textrm{H}$ changes its sign.
In K$_{0.49}$RhO$_2$, no anomalies are detected even in $dS/dT$ around 50~K, which implies that the two-carrier model is unlikely.
Third possibility is that $T$-dependent $R_\textrm{H}$ in K$_{0.49}$RhO$_2$ is caused by a special geometry of the crystal structure.
In fact, $T$-dependent $R_\textrm{H}$ of triangular- and kagom\'e-lattice systems is theoretically suggested by means of high-temperature expansion formulae\cite{rh_t1,rh_t2,rh_t3} and consistent with experiment\cite{NCO_hall}.
Koshibae and Maekawa suggest that the electronic states in triangular-lattice system of the Co oxides are effectively regarded as four interpenetrating kagom\'e lattices\cite{kagome}.
From the structural similarities, layered Rh oxides are expected to show such behavior.
The high-temperature $R_\textrm{H}$ of these systems depends on temperature as\cite{rh_t3} 
\begin{equation}
 R_\textrm{H}(T)=-\frac{v}{4e}\frac{k_\textrm{B}T}{t}\frac{1+y}{y(1-y)},
 \label{eq3}
\end{equation}
where $v$, $t$, and $y$ are the unit cell volume, the transfer integral, and the electron concentration, respectively.
This indicates that $R_\textrm{H}$ is linear in $T$.

\begin{figure}[t]
  \begin{center}
   \includegraphics[width=8cm,clip]{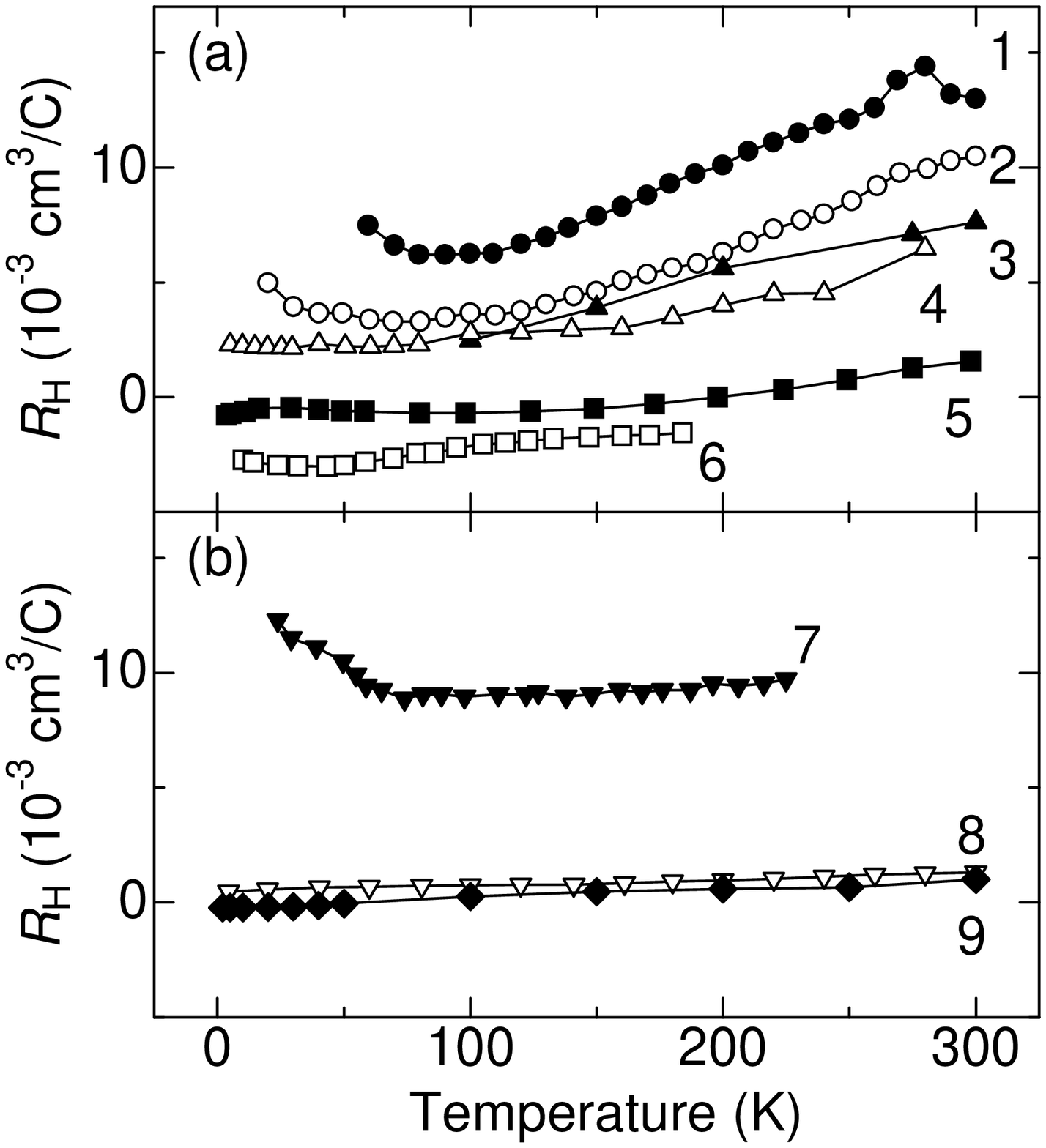}
   \caption{
 Hall coefficient of various kind of the layered (a) Co and (b) Rh oxides.
 The numbers indicate 1: Ca$_{2.8}$Nd$_{0.2}$Co$_4$O$_9$\cite{CNCO_hall}, 2: Bi$_{1.56}$Pb$_{0.44}$Sr$_2$Co$_{1.9}$O$_y$\cite{BSCO_hall}, 3: Ca$_3$Co$_4$O$_9$\cite{CCO_hall}, 
 4: $[$Sr$_2$O$_{2-\delta}]_{0.53}$CoO$_2$\cite{SCO_hall}, 5: Na$_{0.68}$CoO$_2$\cite{NCO_hall}, 6: NaCo$_{1.9}$Pd$_{0.1}$O$_4$\cite{NCPO},
 7: (Bi$_{0.8}$Pb$_{0.2}$)$_{1.8}$Ba$_2$Rh$_{1.9}$O$_y$\cite{BBRO}, 8: Bi$_{0.78}$Sr$_{0.4}$RhO$_{3+\delta}$\cite{BSRO_single}, and 9: K$_{0.49}$RhO$_2$.
   }
   \label{fig3}
  \end{center}
\end{figure}

Let us apply eq.~(\ref{eq3}) to K$_{0.49}$RhO$_2$ according to the previous study in Bi$_{0.78}$Sr$_{0.4}$RhO$_{3+\delta}$\cite{BSRO_single}.
Figure~\ref{fig3} shows $T$-dependence of $R_\textrm{H}$ of various layered Co/Rh oxides.
We need to express that the Co/Rh ions are stable in the low-spin state and the electronic structure should be same in the layered Co/Rh oxides.
We can see the $T$-linear $R_\textrm{H}$ in all materials, and find a clear difference in $dR_\textrm{H}/dT$ between Co (Nos. 1 - 6 in Fig.~\ref{fig3}) and Rh oxides (Nos. 7 - 9 in Fig.~\ref{fig3}).
In addition to these materials, TiS$_2$ with triangular lattice also shows $T$-linear $R_\textrm{H}$\cite{TiS2}, which seems qualitatively consistent with eq.~(\ref{eq3}).
Table~\ref{tab1} summarizes $dR_\textrm{H}/dT$ of these layered Co/Rh oxides in the unit of 10$^{-6}$~cm$^3$/C$\cdot$K from $T$-linear fit above 100~K.
Eq.~(\ref{eq3}) indicates that $dR_\textrm{H}/dT$ dependes on $t$ and carrier concentration.
In real materials such as Ca-Co-O and Bi-Sr-Co-O, the $T$-linear part of $R_\textrm{H}$ has weak relation with carrier concentration\cite{CNCO_hall,BSCO_hall}.

To see the relation between the thermopower and $R_\textrm{H}$, we also put thermopower at 300~K in Table~\ref{tab1}.
The Rh oxides show smaller $dR_\textrm{H}/dT$ and thermopower than the Co oxides\cite{Rhmisfit}, which may be due to the difference between $3d$ and $4d$ orbitals.
The misfit oxides (Nos. 1, 2, 3, 4, and 7) have $T$-independent thermopower at 300~K, which has been explained by the Heikes formula.
Since the Heikes formula is independent of the band width $t$, $dR_\textrm{H}/dT$ has little to do with the magnitude of $S(T)$.

Equation~(\ref{eq3}) shows that the carrier concentration dependence of $R_\textrm{H}$ is almost independent of $y$ where the carrier concentration is nearly half ($y\sim 0.5$), 
and accordingly the four-times difference in $dR_\textrm{H}/dT$ between Na$_{0.68}$CoO$_2$ and K$_{0.49}$RhO$_2$ is ascribed to the difference in $t$.
In fact, the estimated $t$'s from these $dR_\textrm{H}/dT$ in Na$_{0.68}$CoO$_2$ and K$_{0.49}$RhO$_2$ are $\sim$25~K\cite{NCO_hall} and $\sim$160~K, respectively.
Kobayashi \textit{et al.} report $t\sim$200~K in Bi$_{0.78}$Sr$_{0.4}$RhO$_{3+\delta}$\cite{BSRO_single}.
We should note that angle-resolved photoemission spectroscopy data of Na$_{0.7}$CoO$_2$ suggest $t\sim$120~K\cite{NCO_arpes2} which is different from the evaluated $t\sim$25~K for Na$_{0.68}$CoO$_2$\cite{NCO_hall}.
This means that $t$ evaluated from $R_\textrm{H}$ may scale the band width of the material, but is not strictly equal.
Nevertheless, we can see that Co oxides are more correlated than Rh oxides.
We should note that eq.~(\ref{eq3}) is valid for infinitely large Coulomb repulsion, which is not realized in real materials.
Furthermore, $T$-linear $R_\textrm{H}$ is derived from the assumption that the material of $y\sim0$ is a Mott insulator, but the real material (\textit{i.e.} CoO$_2$) is a paramagnetic metal\cite{CoO2}.
Further studies of both experimentally and theoretically are necessary.

\begin{table}[t]
 \caption{Temperature coefficient of $R_\textrm{H}$ of various layered Co/Rh oxides above 100~K.
 There is a clear difference in the slope of $R_\textrm{H}$ between Co and Rh.}
 \begin{tabular}{cccccc}
  \hline \hline
  No. & Material & Slope ($10^{-6}$~cm$^3$/C$\cdot$K) & $S_\textrm{\scriptsize{300~K}}$ ($\mu$V/K) & Crystal & Ref. \\
  \hline
  1 & Ca$_{2.8}$Nd$_{0.2}$Co$_4$O$_9$ & 44 & 147 & poly & \cite{CNCO_hall} \\
  2 & Bi$_{1.56}$Pb$_{0.44}$Sr$_2$Co$_{1.9}$O$_y$ & 32 & --- & single & \cite{BSCO_hall} \\
  3 & Ca$_3$Co$_4$O$_9$ & 31 & 129 & single & \cite{CCO_hall} \\
  4 & $[$Sr$_2$O$_{2-\delta}]_{0.53}$CoO$_2$ & 14 & 70 & single & \cite{SCO_hall} \\
  5 & Na$_{0.68}$CoO$_2$ & 11 & 90 & single & \cite{NCO_hall} \\
  6 & NaCo$_{1.9}$Pd$_{0.1}$O$_4$ & 6$^\dagger$ & 35 & poly & \cite{NCPO} \\
  7 & (Bi$_{0.8}$Pb$_{0.2}$)$_{1.8}$Ba$_2$Rh$_{1.9}$O$_y$ & 4.4 & 95 & poly & \cite{BBRO} \\
  8 & Bi$_{0.78}$Sr$_{0.4}$RhO$_{3+\delta}$ & 2.7 & 65 & single & \cite{BSRO_single} \\
  9 & K$_{0.49}$RhO$_2$ & 2.7 & 40 & single & This work \\
  \hline \hline
 \end{tabular}

 $^\dagger$ Data exist only below 200~K.
 \label{tab1}
\end{table}

\section{Summary}
We have measured the transport properties of single-crystalline samples of K$_{0.49}$RhO$_2$ which is isomorphic to $\gamma$-Na$_x$CoO$_2$.
Resistivity and thermopower at 300~K are 400~$\mu\Omega$cm and 40~$\mu$V/K, respectively.
Resistivity shows $T^2$-dependence up to 300~K and thermopower is proportional to temperature up to 140~K.
Temperature dependence of transport properties of K$_{0.49}$RhO$_2$ are different from those of Na$_x$CoO$_2$, 
which suggests that the correlation in Rh oxides are weaker than that of Na$_x$CoO$_2$.
The estimated power factor at 300~K is 4.0~$\mu$W/cmK$^2$ which is the largest among other layered Rh oxides.
The Hall coefficient is almost proportional to temperature above 100~K, which is similar to other layered Co/Rh oxides.
Quantitative analysis of this $T$-linear term of Hall coefficient indicates that the effective band width related to the magnetotransport is much larger in the Rh oxides.
Further study is needed for the understanding of Co/Rh oxides.

\begin{acknowledgments}
We thank W. Kobayashi, Y. Klein, M. Abdel-Jawad, K. Kuroki, T. Katsufuji, and D.J. Singh for fruitful discussions.
This work was partially supported by a Grant-in-Aid for JSPS Fellows.
\end{acknowledgments}

\end{document}